\documentclass{ns07abs}
\usepackage{amsmath}
\usepackage{cite}
\usepackage{graphicx}
\usepackage[figuresright]{rotating}
\usepackage{geometry}
\newcommand{\bra}[1]{\left\langle #1 \right|}

\newcommand{\ket}[1]{\left| #1 \right\rangle}

\newcommand{\Hhat}{\hat{H}}

\newcommand{\Ihat}{\hat{I}}
\newcommand{\Jhat}{\hat{J}}

\newcommand{\Hc}{{\cal H}}

\newcommand{\Jc}{{\cal J}}

\newcommand{\Nhat}{\hat{N}}

\newcommand{\Qhat}{\hat{Q}}

\newcommand{\Phat}{\hat{P}}

\newcommand{\Psihat}{\hat{\Psi}}

\newcommand{\del}{\partial}

\newcommand{\beq}{\begin{equation}}
\newcommand{\beqa}{\begin{eqnarray}}
\newcommand{\eeq}{\end{equation}}
\newcommand{\eeqa}{\end{eqnarray}}

\begin{document}

\title{MICROSCOPIC DYNAMICS OF SHAPE COEXISTENCE PHENOMENA AROUND
 $^{68}$Se AND $^{72}$Kr}
\author{NOBUO HINOHARA$^{*}$, KENICHI MATSUYANAGI}
\address{Department of Physics, Graduate School of Science, Kyoto
University, \\
 Kyoto 606-8502, Japan \\
$^*$E-mail: hinohara@ruby.scphys.kyoto-u.ac.jp}
\author{TAKASHI NAKATSUKASA}
\address{Theoretical Nuclear Physics Laboratory, RIKEN Nishina Center, \\
Wako 351-0198, Japan}
\author{MASAYUKI MATSUO}
\address{Department of Physics, Faculty of Science, Niigata University,
\\
Niigata 950-2181, Japan}
\maketitle
\begin{abstract}
The adiabatic self--consistent collective coordinate (ASCC) method is 
applied to the pairing-plus-quadrupole (P + Q) model Hamiltonian including 
the quadrupole pairing, and the oblate--prolate shape coexistence phenomena 
in proton-rich nuclei, $^{68}$Se and $^{72}$Kr, are investigated.
It is shown that the collective path connecting the oblate and prolate 
local minima runs along a triaxial valley in the $(\beta,\gamma)$ plane.
Quantum collective Hamiltonian is constructed 
and low-lying energy spectra and E2 transition probabilities 
are calculated for the first time using the ASCC method.
Basic properties of the shape coexistence/mixing are well reproduced.
We also clarify the effects of the time-odd pair field on the collective mass 
(inertial function) for the large--amplitude vibration and on the rotational moments 
of inertia about three principal axes.
\end{abstract}



\section{Introduction}
In proton rich nuclei around $^{68}$Se and $^{72}$Kr, 
oblate--prolate shape coexistence phenomena are observed
\cite{fis00,fis03,bou03,gad05}.
From the mean-field point of view, the oblate and prolate shapes are mixed
by the many-body tunneling through the potential 
barrier between the two local minima in the potential energy landscape.
To describe the shape mixing dynamics, 
we have to determine the collective degrees of freedom in which
direction the large--amplitude shape dynamics takes place.
The adiabatic self--consistent collective coordinate (ASCC) method
\cite{mat00}
has been proposed as a microscopic theory of large--amplitude collective
motion such as the shape coexistence phenomena.
This theory is based on the time-dependent Hartree--Fock--Bogoliubov
(TDHFB)
theory and enables us to extract the collective degrees of freedom
in a self--consistent way.
In the previous work \cite{kob05}, we have solved the 
pairing-plus-quadrupole (P + Q) Hamiltonian by means of the ASCC method 
for $^{68}$Se and $^{72}$Kr nuclei, and successfully extracted 
the one-dimensional collective path 
connecting the oblate and prolate local minima.
Since the two local minima are mainly connected by the triaxial degrees of
freedom, we have suggested that the triaxial deformation plays 
an essential role in the shape coexistence dynamics of these nuclei.

In this presentation, we report results of the first calculation of 
low-lying energy spectra and E2 transition probabilities 
by means of the ASCC method. We derive the quantum collective Hamiltonian 
that describes the coupled collective motion of 
the large--amplitude vibration responsible for the oblate--prolate shape mixing 
and the three-dimensional rotation of the triaxial shape. 
The calculation yields the excited prolate rotational band as well as the 
oblate ground-state band. It also indicates that the 
shape mixing decreases as the angular momentum increases. 

We also clarify the effect of the time-odd mean-field on the collective mass. 
This effect is ignored in the cranking mass~\cite{rin80}, but is included 
in the ASCC mass. The time-odd mean-field effect generated by the 
particle-hole residual interaction was investigated in a few decades
ago~\cite{dob81}, but those generated by the pairing interaction
has not been discussed so far. Quite recently we have shown, using the 
schematic model Hamiltonian~\cite{hin06}, that the time-odd component 
associated with the quadrupole-type pair field significantly increases 
the collective mass. In the present calculation,  we thus 
include the quadrupole pairing interaction to the P + Q Hamiltonian 
to clarify the effect of the time-odd pair field on the collective 
mass and rotational moments of inertia.

\section{The ASCC method}

We first recapitulate the basic equations of the ASCC method.
The moving-frame TDHFB state 
$\ket{\phi(q,p)} = e^{ip\Qhat(q)}\ket{\phi(q)}$
is written in terms of the collective coordinate $q$,
and collective momentum $p$.
The TDHFB state is expanded in terms of the
collective momentum $p$ under the adiabatic assumption.
The collective Hamiltonian is expanded up to the lowest order in $p$ as
\begin{align}
 \Hc(q,p,\vec{I}) = V(q) + \frac{1}{2}B(q)p^2 
+\sum_{i=1}^3 \frac{I_i^2}{2\Jc_i(q)}, \label{eq:collH}
\end{align}
where the $V(q)$ and $B(q)$ represents the 
collective potential and the inverse collective mass.
We add the rotational energy term with three moments of inertia
$\Jc_i(q)$ to the collective Hamiltonian in 
order to take into account the rotational motion.

The basic equations of the ASCC method are derived from the adiabatic 
expansion of the time-dependent variational principle, and are
summarized as
\begin{align}
\delta \bra{\phi(q)}\Hhat_M(q)\ket{\phi(q)} = 0, \label{eq:mfHFB}
\end{align}
\begin{align}
\delta
 \bra{\phi(q)}[\Hhat_M(q),\Qhat(q)]-\frac{1}{i}B(q)\Phat(q)\ket{\phi(q)}
= 0, \label{eq:mfQRPA1}
\end{align}
\begin{multline}
\delta  \bra{\phi(q)}[\Hhat_M(q),\frac{1}{i}\Phat(q)] -C(q)\Qhat(q) \\
-\frac{1}{2B(q)}[[\Hhat_M(q),\frac{dV}{dq}(q)\Qhat(q)], \Qhat(q)]
- \sum_{\tau} \frac{\del \lambda^{(\tau)}}{\del
	     q}(q)\Nhat^{(\tau)}\ket{\phi(q)} = 0.
 \label{eq:mfQRPA2}
\end{multline}
Equation (\ref{eq:mfHFB}) is called the moving-frame HFB equation, while 
Eqs. (\ref{eq:mfQRPA1}) and (\ref{eq:mfQRPA2}) are called the
moving-frame QRPA equations.
where $\Qhat(q)$ and $\Phat(q)$ denote the infinitesimal generators of
the collective path,
\begin{align}
 \Hhat_M(q) = \Hhat - \sum_{\tau} \lambda^{(\tau)}(q)\Nhat^{(\tau)}
- \frac{dV}{dq}(q)\Qhat(q),
\end{align}
is the Hamiltonian in the moving-frame, and
$C(q) = \frac{\del^2 V}{\del q^2} + \frac{1}{2B(q)}\frac{\del B}{\del q}
\frac{\del V}{\del q}$ 
is the local stiffness parameter. The operators $\Hhat$,
$\Nhat^{(\tau)}$
and the quantity $\lambda^{(\tau)}(q)$ represents 
the microscopic Hamiltonian, the particle number operators,
and the chemical potential, respectively.
The QRPA eigenmodes about the HFB equilibrium points, 
as well as static solutions of the HFB equation, are 
always special solutions of the ASCC equations. 
Therefore, the collective path can be constructed 
by means of the local progression procedure starting from a static HFB state.
The collective path at $q+\delta q$ is calculated using the constraint
$\bra{\phi(q+\delta q)}\Qhat(q)\ket{\phi(q+\delta q)} = \delta q$
derived from the canonical variable conditions,
if the collective path at $q$ is already known.
Repeating this procedure, we can find the collective path and obtain 
all quantities appearing in the ASCC equations and collective
Hamiltonian except the rotational moments of inertia.
Three rotational moments of inertia are evaluated by extending the QRPA equations 
for rotation at the HFB minima to the general HFB states 
$\ket{\phi(q)}$ on the collective path,
\begin{gather}
 \delta \bra{\phi(q)} [\Hhat_M(q), i\Psihat_i(q)] - \Jc_i^{-1}(q)
 \Jhat_i\ket{\phi(q)}= 0,
\end{gather}
where $\Psihat_i(q)$ and $\Jhat_i$ are the angle and angular momentum
operators in the principal frame.

In the present calculation, we use the P + Q + quadrupole pairing 
Hamiltonian as a microscopic Hamiltonian $\Hhat$. 
We adopt the same single-particle energies and the P + Q interaction strengths 
as in Ref.~\citen{kob05} and the self--consistent quadrupole-pairing 
strength $G_2^{\rm self}$ proposed by Sakamoto and Kishimoto~\cite{sak90}.

\section{Collective Path}

\begin{figure}[ht]
\begin{center}
\begin{tabular}{c}
\hspace*{40mm}\includegraphics[width=100mm]{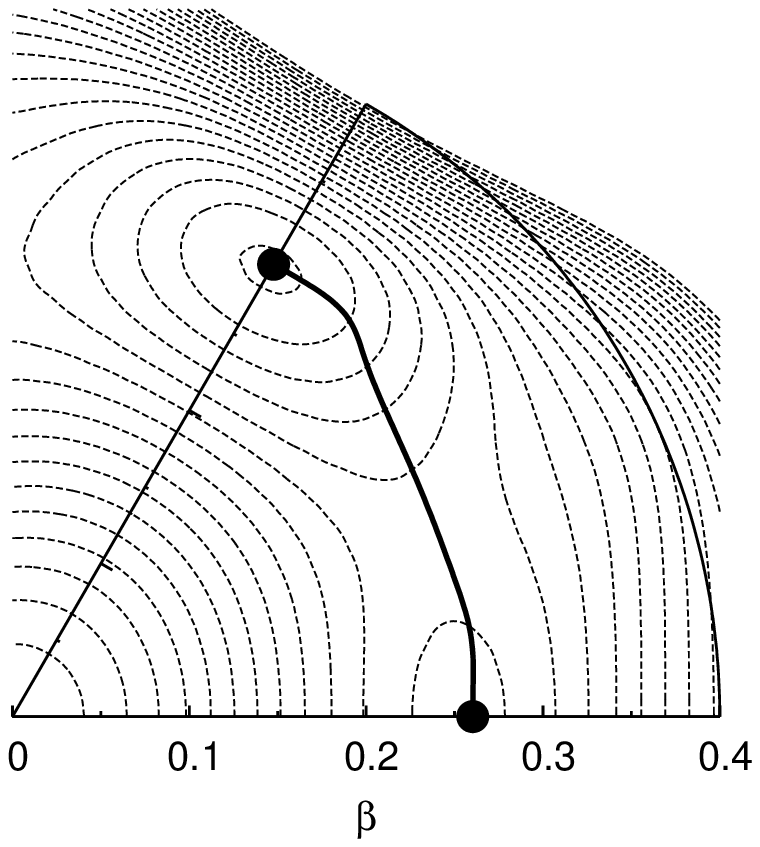} \\
\hspace*{40mm}\includegraphics[width=100mm]{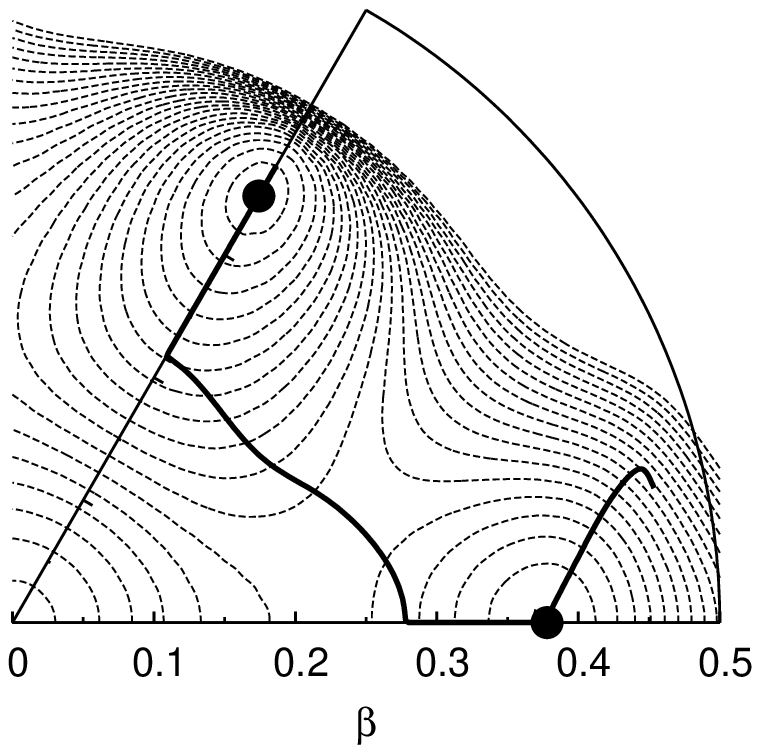}
\end{tabular}
\caption{
Collective paths for $^{68}$Se ({\it upper}) and $^{72}$Kr ({\it lower})
calculated using the P + Q Hamiltonian including the
quadrupole pairing interaction.
The paths are projected onto the $(\beta,\gamma)$ potential
energy surface. The dots in the figures indicate the HB local minima.
The equipotential contour lines are drawn every 100 keV.
}
\label{fig:path}
\end{center}
\end{figure}

\begin{figure}[ht]
\begin{center}
\begin{tabular}{c}
\includegraphics[width=87mm]{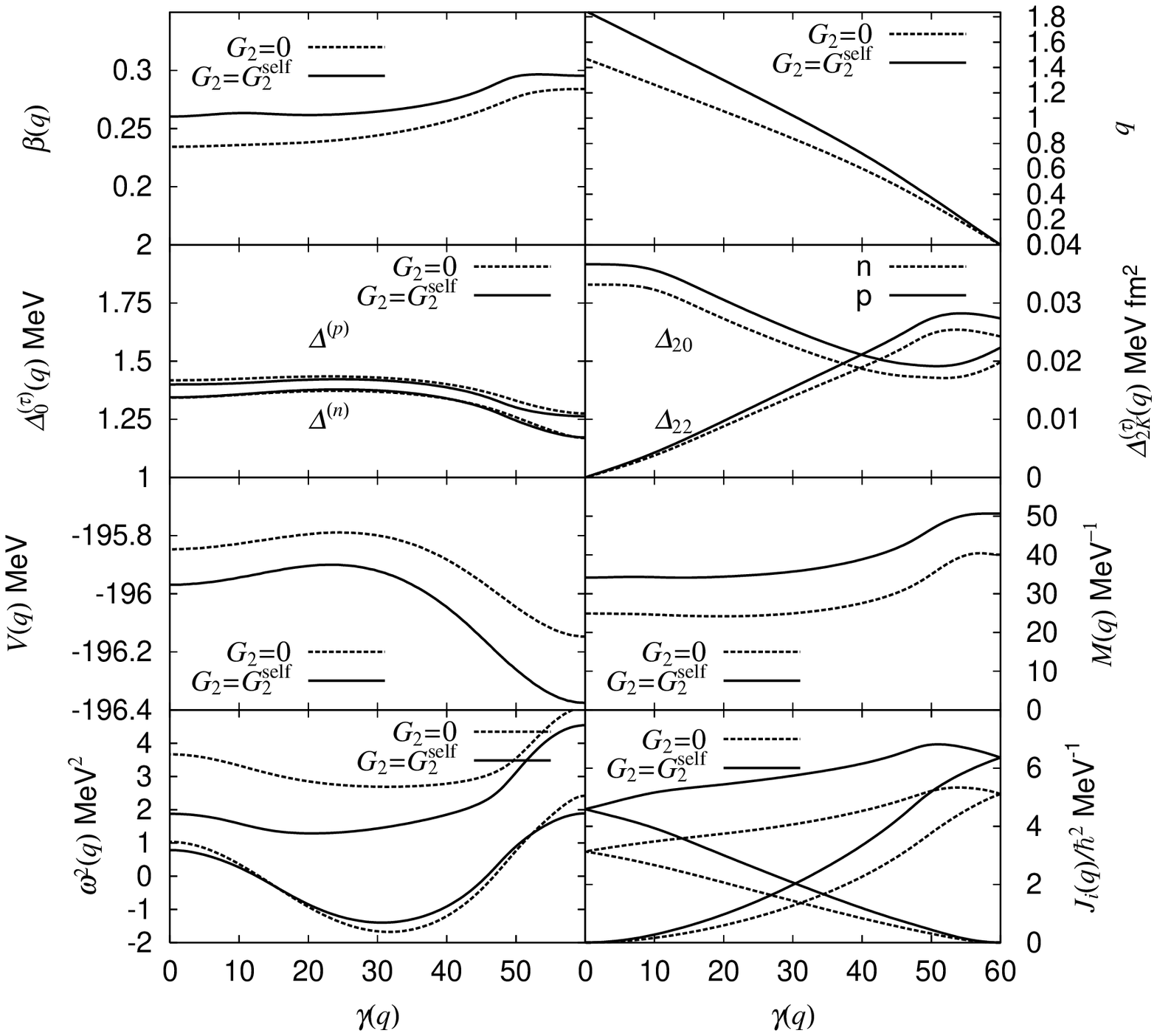} \\
\includegraphics[width=87mm]{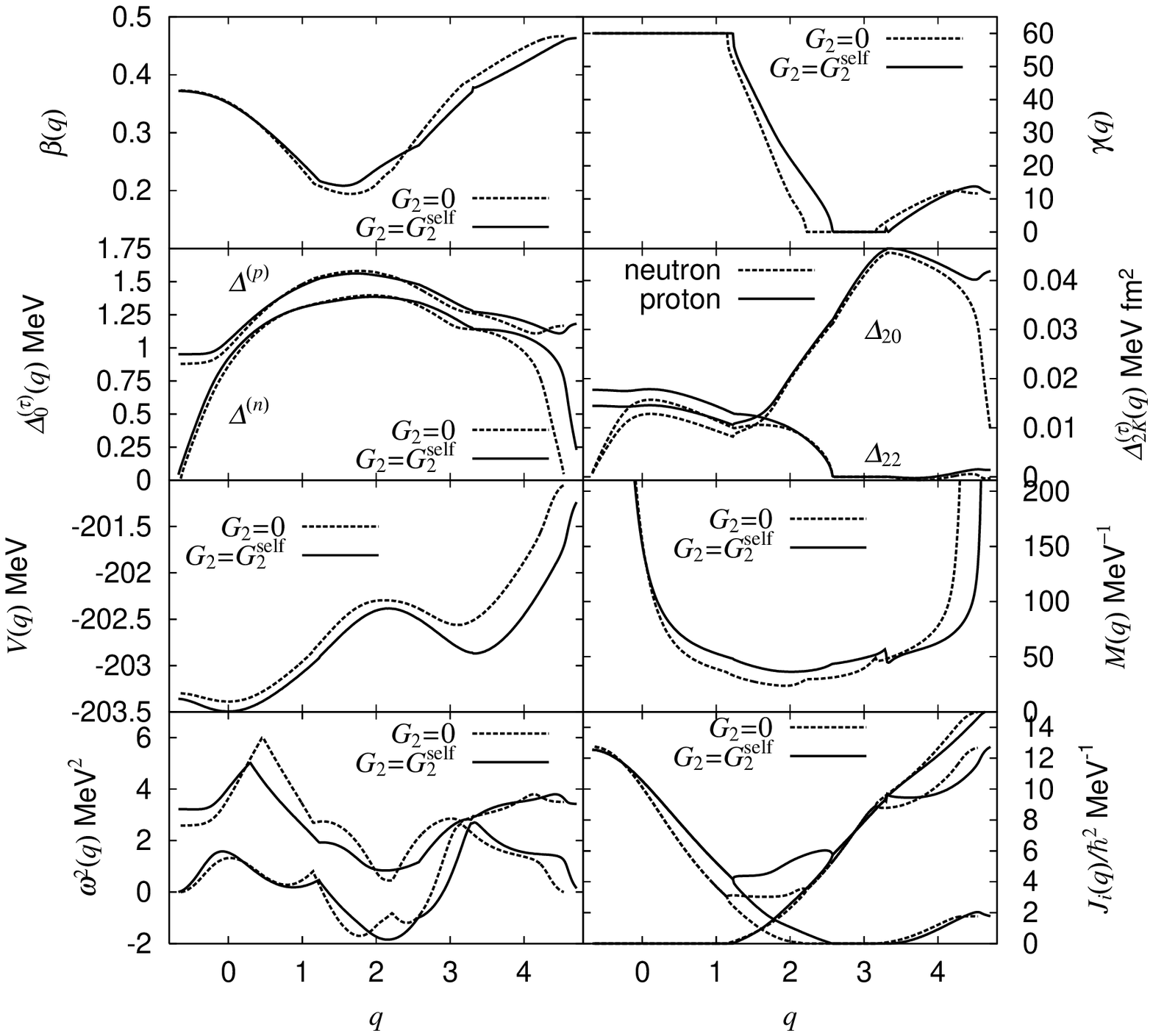}
\end{tabular}
\caption{The monopole pairing gaps, $\Delta_{0}^{(\tau)}(q)$,
the quadrupole pairing gaps, $\Delta_{20}^{(\tau)}(q)$ and
$\Delta_{22}^{(\tau)}(q)$,
the collective potential $V(q)$, the collective mass $M(q)$,
the rotational moments of inertia $\Jc_i(q)$,
the lowest two moving-frame QRPA frequencies squared,
 $\omega^2(q)=B(q)C(q)$,
and the quadrupole deformations, $\beta(q)$ and $\gamma(q)$, are
plotted as functions of $\gamma(q)$
for $^{68}$Se ({\it upper}) and $q$ for $^{72}$Kr ({\it lower}).
Results of two calculations using the P + Q Hamiltonian with
($G_2=G_2^{\rm self}$) and without $(G_2=0)$ the quadrupole pairing
interaction are compared.}
\label{fig:quantity}
\end{center}
\end{figure}

For both $^{68}$Se and $^{72}$Kr, we have found that 
the lowest HFB state possesses the oblate shape, 
while second lowest HFB state the prolate shape. 
We start from the oblate state ($q=0$) 
and determine the collective path connecting the two local minima.
Figure~\ref{fig:path} shows the collective path projected onto the
$(\beta,\gamma)$ potential energy surfaces. The path connects the two local
minima via the triaxially deformed region. 
Figure~\ref{fig:quantity} shows various quantities appearing 
in the collective Hamiltonian and the basic equations of the ASCC method.
We define the scale of collective coordinate $q$ by setting $B(q)^{-1}=1$. 
The collective mass with respect to the conventional  
($\beta, \gamma$)- coordinates is then given by 
$M(q) = \sqrt{(d\beta/dq)^2+\beta^2(d\gamma/dq)^2}^{-1}$.
We have found that the collective mass $M(q)$ and three rotational moments of 
inertia $\Jc_i(q)$ are enhanced by the time-odd pair field 
generated by the quadrupole pairing.

\section{Energy Spectrum and Transition Probabilities}

We requantize the collective Hamiltonian (\ref{eq:collH}) and 
solve the collective Schr\"odinger equation
\begin{align}
 \left(
 -\frac{1}{2}\frac{\del^2}{\del q^2} + \frac{1}{2}\sum_{i=1}^3 \Jc_i^{-1}(q)\Ihat_i^2
 + V(q)
\right) \Psi_{IM,k}(q,\Omega) = E_{I,k} \Psi_{IM,k}(q,\Omega),
\end{align}
to obtain the excitation energies and collective wavefunctions.
The collective wave functions $\Psi_{IM,k}(q,\Omega)$ takes the following form: 
\begin{align}
 \Psi_{IM,k}(q,\Omega) = \sum_{K=0}^I \Phi_{IK,k}(q) <\Omega|IMK>,
\end{align}
where $\Phi_{IK,k}(q)$ and $<\Omega|IMK>$ denote the vibrational 
wavefunctions and the rotation matrices, respectively.
The conventional boundary conditions and symmetry requirements 
for solving the Bohr collective Hamiltonian~\cite{kum67} are adopted.

\begin{figure}[ht]
\begin{center}
\begin{tabular}{c}
\includegraphics[width=120mm]{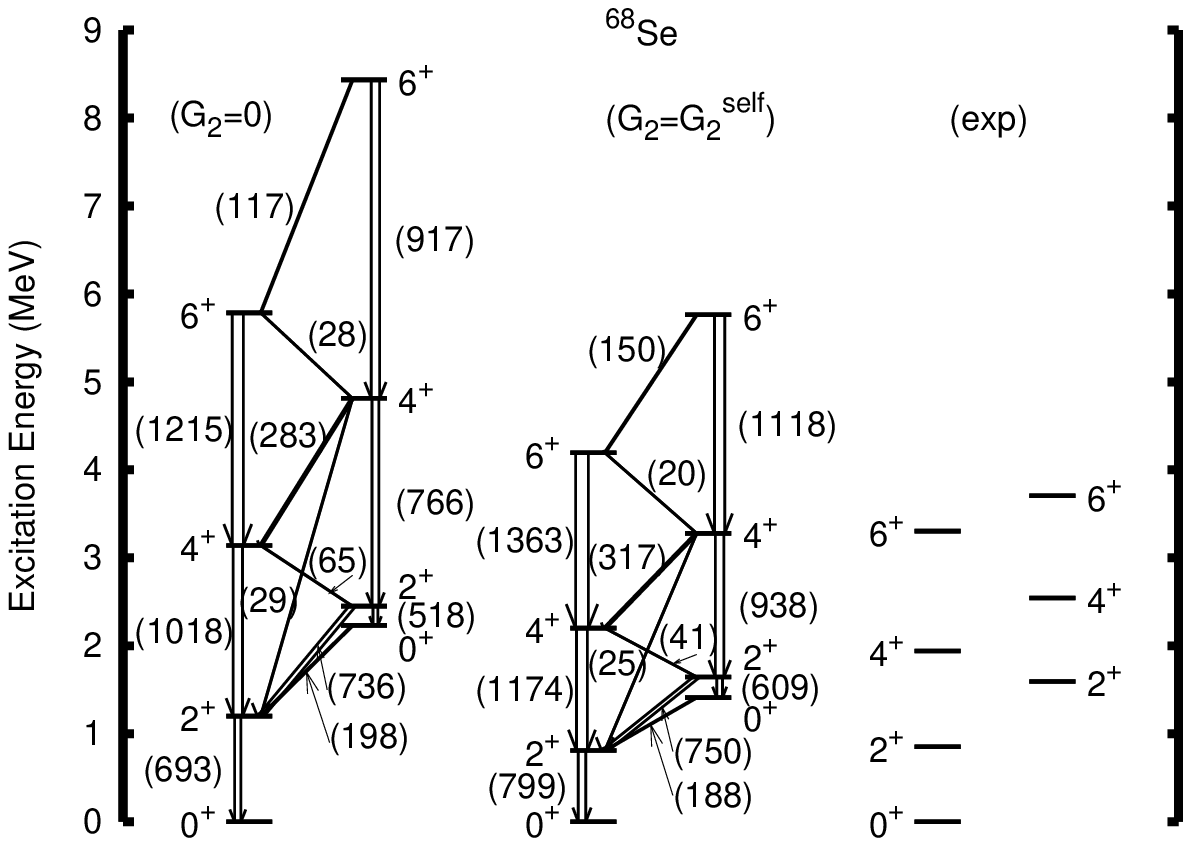} \\
\includegraphics[width=120mm]{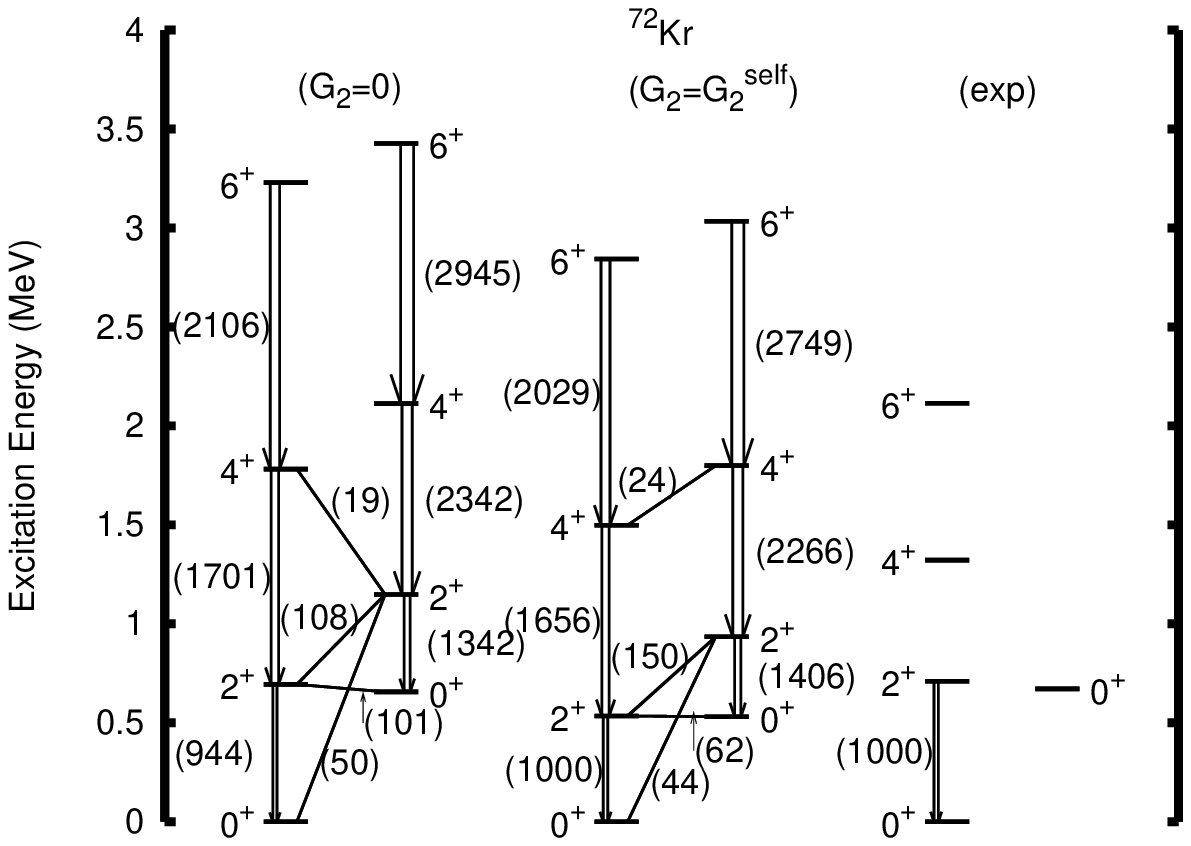}
\end{tabular}
\end{center}
\caption{Excitation energies and $B$(E2) values for low-lying states of
 $^{68}$Se
and $^{72}$Kr calculated by the ASCC method.
In the left (middle) panel, the quadrupole pairing is ignored
 (included).
Experimental data~\cite{fis00,fis03,bou03,gad05} are displayed in the
 right panel.
The $B$(E2) values are given in parentheses beside the arrows
in unit of $e^2$ fm$^4$.}
\label{fig:68Se-energy}
\end{figure}

Figures~\ref{fig:68Se-energy} display  
the energy spectrum and $B$(E2) values for $^{68}$Se and $^{72}$Kr.
The calculation yields the excited prolate rotational band as well as the 
oblate ground-state band. It is seen that the inter-band E2 transitions 
are weaker than the intra-band E2 transitions, indicating that 
the oblate--prolate shape coexistence picture holds. 
The result of calculation also indicates that the oblate--prolate shape mixing 
decreases as the angular momentum increases. 
Thus, the basic patterns of oblate--prolate shape coexistence 
are successfully reproduced in the calculation.
The calculation suggest the existence of excited $0^+$ state 
in $^{68}$Se, which is not yet found in experiment.
By comparing two calculations, ($G_2=0$) and ($G_2=G_2^{\rm self}$),
we found that the energy of the $0_2^+$ state is much lowered by including the 
quadrupole pairing interaction. This is due to the enhancement of the
collective mass. The energies of the other members of the rotational bands 
are also lowered, because the quadrupole pairing also enhances the rotational moments of 
inertia. 
Experimental data for $B$(E2) are not available except 
$B($E2$;2_1^+\rightarrow 0_1^+$)=1000 $e^2$ fm$^4$ in $^{72}$Kr~\cite{gad05}.
Theoretical $B$(E2) values are calculated using a polarization charge
$e_{\rm pol}=0.881$ 
that reproduces this data,
so that only relative magnitudes should be regarded as theoretical estimates.

\begin{figure}[ht]
\begin{center}
\begin{tabular}{cc}
\includegraphics[width=65mm]{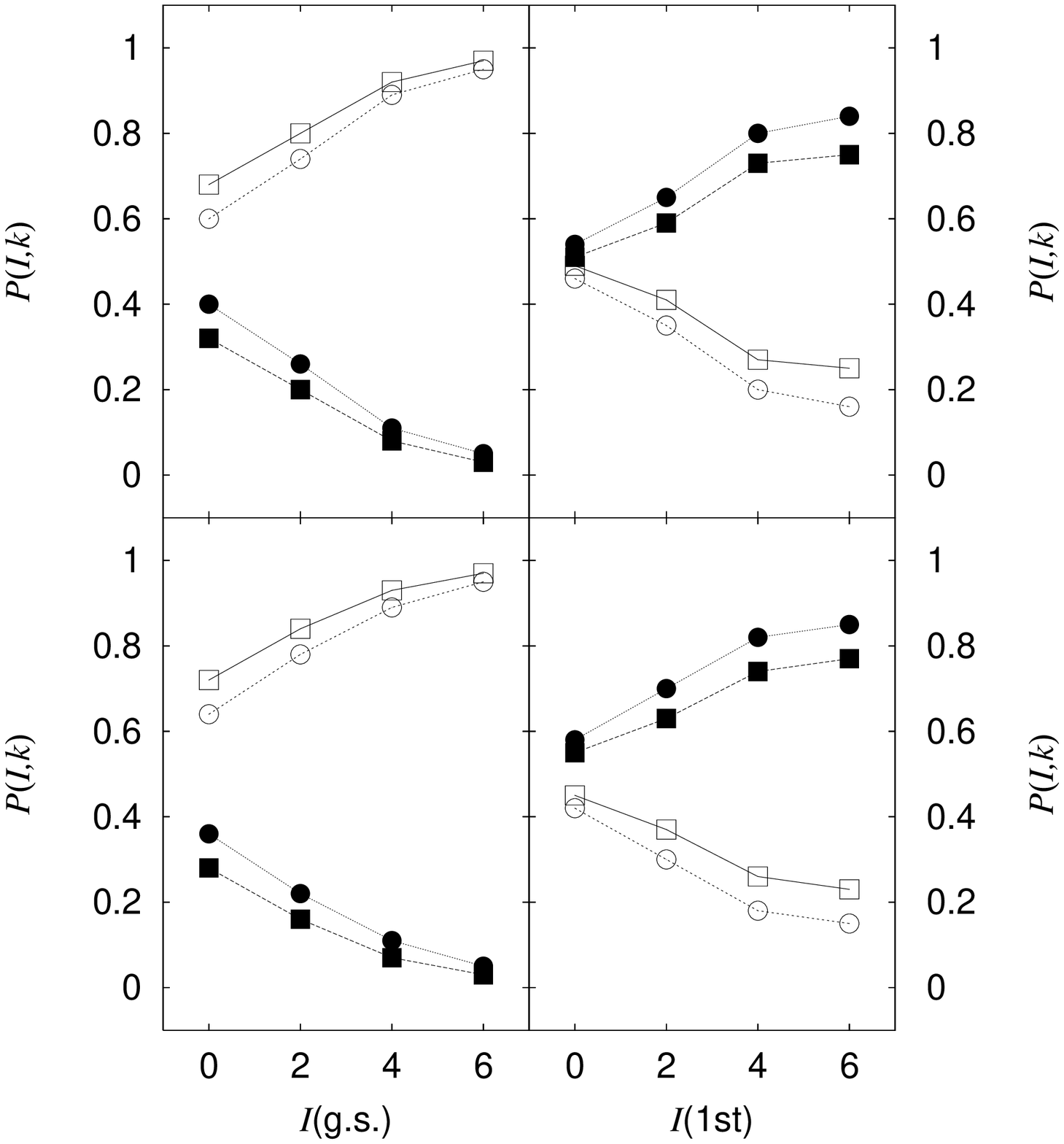} &
\includegraphics[width=65mm]{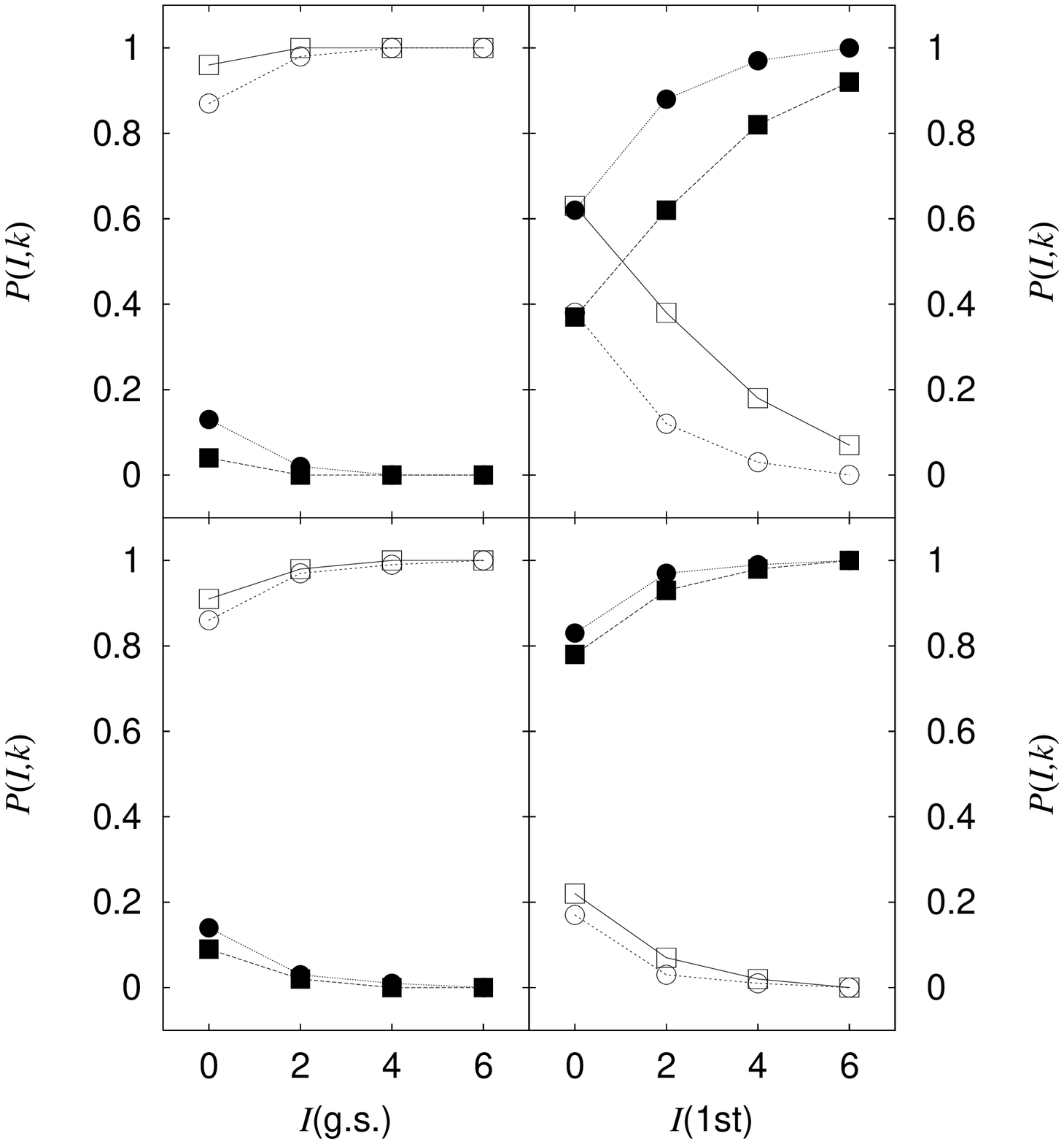}
\end{tabular}
\end{center}
\caption{
The oblate and prolate probabilities evaluated for individual
eigenstates in $^{68}$Se ({\it left}) and $^{72}$Kr ({\it right}).
For each nucleus, the left (right) panel shows values for the lowest
(the second lowest) state of each angular momentum.
The open (closed) symbols indicate the oblate (prolate) probabilities.
The probabilities defined by setting the boundary at the barrier top
($\gamma=30^\circ$) are shown by squares (circles)
In each figure, the upper (lower) panel shows the probabilities
calculated without (with) including the quadrupole pairing interaction.}
\label{fig:mixing}
\end{figure}

In order to estimate the oblate--prolate shape mixing 
in quantum eigenstates, we define the oblate and prolate 
probabilities as follows: 
\begin{align}
 P_{\rm ob}(I,k) = \int_{q_{\rm min}}^{q_0} dq 
\sum_{K=0}^I |\Phi_{IK,k}(q)|^2, \quad
 P_{\rm pro}(I,k) = \int_{q_0}^{q_{\rm max}} dq \sum_{K=0}^I
 |\Phi_{IK,k}(q)|^2,
\end{align}
where we assume that 
$q_{\rm min} \le q_{\rm ob} < q_0 < q_{\rm pro} \le q_{\rm max}$.
The ``boundary'' between the oblate and the prolate regions is set to 
the top of the potential barrier or at $\gamma=30^\circ$.
Figure~\ref{fig:mixing} shows these probabilities for $^{68}$Se and
$^{72}$Kr.
It is clearly seen that the shape mixing rapidly decreases 
as the angular momentum increases.

\section{Summary}
For the first time, we have reported the result of calculation of 
low-lying energy spectra and E2 transition probabilities in
$^{68}$Se and $^{72}$Kr by means of the ASCC method. 
We have derived the quantum collective Hamiltonian that describes 
the coupled collective motion of the large--amplitude vibration 
responsible for the oblate--prolate shape mixing 
and the three-dimensional rotation of the triaxial shape. 
The calculation yields the excited prolate rotational band as well as the 
oblate ground--state band. It also indicates that the oblate--prolate shape 
mixing decreases as the angular momentum increases. 

It is surprising that basic pattern of the shape coexistence/mixing phenomena 
is well reproduced using the one-dimensional collective path running on the 
two-dimensional $(\beta,\gamma)$ plane. Speaking more exactly, this 
collective path is self--consistently extracted from the huge dimensional 
TDHB manifold. Namely, the result of calculation indicates that 
the TDHB collective dynamics of the shape coexistence phenomena 
in these nuclei is essentially controlled by the single collective variable 
microscopically derived by means of the ASCC method.

We have also shown that the time-odd pair field enhances the
collective mass of the large--amplitude vibrational motion and 
the rotational moments of inertia. 
This finding is important in understanding the 
shape coexistence dynamics, because, together with the collective potential 
energy, these inertial functions associated with the collective kinetic 
energies determine the degree of localization of the collective wave function 
in the $(\beta, \gamma)$ plane.

\section*{Acknowledgement}
This work is supported by the Grant-in-Aid for the 21st Century COE
``Center for Diversity and Universality in Physics'' from
the Ministry of Education, Culture, Sports, Science and Technology
(MEXT) of Japan.

\newpage

\newpage

\newpage

\newpage

\end{document}